\documentclass[a4paper]{JHEP3}
\usepackage{epsfig}
\usepackage{graphicx}
\usepackage{amssymb}
\usepackage{latexsym}

\newcommand{\be}{\begin{equation}}
\newcommand{\ee}{\end{equation}}
\newcommand{\bea}{\begin{eqnarray}}
\newcommand{\eea}{\end{eqnarray}}

\title{Continuum limit of string formation in 3-d SU(2) LGT}
\author{N.D. Hari Dass
\\ Hayama Center for Advanced Studies, Hayama, Japan \\
 Email: \email{hari@soken.ac.jp}}
 \author{Pushan Majumdar
 \\ Institut f\"ur Theoretische Physik,
Westf\"alische Wilhelms-Universit\"at M\"unster. \\
Email: \email{pushan@uni-muenster.de}}

\abstract{
We
 study the continuum limit of the string-like behaviour of flux tubes
 formed between static quarks and anti-quarks in three dimensional $SU(2)$ lattice gauge theory.
We compare our simulation data with the predictions of both effective string models 
as well as perturbation theory. On the string side we obtain clear evidence for
convergence of data to predictions of Nambu-Goto theory. We comment on the scales
at which the static potential starts 
departing from one loop perturbation theory and then again being well 
described by effective string theories. We
 also estimate the leading corrections to the one-loop
  perturbative potential as well as the  Nambu-Goto effective string.
  In the intermediate regions we find that a modified Lennard-Jones type
potential gives surprisingly good fits.
}


\preprint{MS-TP-07-2}

\begin{document}

\section{Introduction}

Gluonic dynamics at large distances can be very effectively probed by looking at properties of the 
flux tube which forms between a static quark and an anti-quark in the QCD vacuum.
Although there is still no analytical proof of this formation, lattice simulation 
results overwhelmingly indicate that this indeed is the case \cite{bali}.

Bosonic string descriptions of this flux tube have been around for a long time.
While most of the earlier attempts \cite{God} used an open string description with 
the ends of the string ending on the quark and the anti-quark, recently there have 
been attempts to give closed string descriptions too.
After Polchinski and Strominger(PS) \cite{PS} suggested how effective theory of strings
 with vanishing conformal anomaly could be formulated
 in all dimensions it has been shown \cite{drum,pmd,unpub} that the
spectrum of these effective theories is universal\footnote{In this article universal will 
mean independent of the details of the underlying gauge group.}
to order $r^{-3}$ ($r$ being the length
of the string) and that to this order they coincide with the predictions of Nambu-Goto
theory whose conformal anomaly however vanishes only in 26 dimensions.

Another interesting idea put forward by L\"uscher and Weisz \cite{lweisz}
is that of open-closed string duality. 
They showed that this too constrained 
the possible string spectra.
In three dimensions they found that this duality implied that to order $r^{-3}$ the spectrum was the 
same as that of the Nambu-Goto(NG) string while in four dimensions one parameter was left 
undetermined. In the PS effective string theories the spectrum to this order is
the same as NG theory in all dimensions without needing to
invoke open-closed duality explicitly.

On the lattice, the flux tube can be observed as the potential between a static quark
 and an anti-quark (open string). This can be obtained from expectation values of Wilson loops or 
 Polyakov loop correlators. One of the characteristics of the string like behaviour of the 
flux tube is the presence of a long distance $1/r$ term in all dimensions and with a 
universal coefficient in the $q\bar q$ potential.  
In four dimensions there is a short distance $1/r\log (r)$ term in the potential. 
However its coefficient depends on the details of the gauge group distinguishing it from
the long distance $1/r$ term. 
The long distance $1/r$ term was first observed in \cite{LSW} and its universality was 
established in \cite{L}. It is known as the L\"uscher term.

The L\"uscher term has been looked at in lattice simulations since the eighties \cite{ambjorn, deF}. 
However 
in recent times, increase in computing power and improvement in algorithms have allowed really 
precise measurements of that term and also the $r^{-3}$ term. 
Now it is possible to study the properties of flux tubes longer than 
1 fermi which was 
unimaginable even a few years back and one is really in a position to compare the 
lattice data with the predictions of the string picture. See \cite{caselle, lw1, lw2, 
 4dim, pushspec1,
 pushspec2, HM5, caslw, kuticr} for example.
In particular, measurements of the L\"uscher term in three dimensions have been 
carried out in \cite{pushspec2, caslw} for $SU(2)$ , in \cite{lw2} for $SU(3)$ and in 
\cite{HM5} for $SU(5)$ lattice 
gauge theories. Convergence
to the static
potential of NG theory to order $r^{-3}$ at a distance scale of around 1 fermi has been 
recently shown in the case
of $d=4$ $SU(3)$ gauge theories in \cite{4dim}.

Another characteristic of the string behaviour of the flux tube is the level spacing
 of the excitation spectrum. We do not discuss it here. See \cite{lweisz, pushspec1, pushspec2,
 Meyer, kutispec} for recent analytic 
 and numerical studies. A review on this topic can also be found in \cite{kutilat05}.

In this article we present results of our simulations of the Polyakov loop
correlators for $d=3$ $SU(2)$ Yang-Mills theory and compare the resulting
static potential with both perturbation theory and string model
predictions. This allows us to narrow down bounds on the distance beyond which we can say the 
flux tube indeed 
shows a string like behaviour.

\section{Simulation parameters}
We have carried out simulations of three dimensional $SU(2)$ lattice gauge theory on 
 lattices at four different lattice spacings with the coarsest lattice having a 
spacing of slightly below 0.13 fm to the finest lattice with a spacing of about 
0.045 fermi. We use symmetric cubic lattices and the Wilson gauge action.
On all these
lattices, we have computed Polyakov loop correlators
$\langle P^*(x)P(y)\rangle$ for various spatial separations $r=y-x$.

To reliably extract signals of these observables which are exponentially
decreasing functions of $r$ and $T$ (the temporal extent of the lattice), 
we used the L\"uscher-Weisz exponential error reduction 
algorithm \cite{lw1}. In this algorithm, one computes intermediate expectation values on 
sub-lattices of the original lattice which are obtained by imposing suitable boundary 
conditions. For measuring Polyakov loop correlators, we obtain our sub-lattices by slicing 
the original lattice along the temporal direction. As is well known, this algorithm has 
several optimization parameters
 among which the number of sub-lattice updates employed seems to be the most important one.
 The lattice parameters along with the number of sub-lattice updates used in 
each set of measurements are summarized in table \ref{tab:run}.

\TABLE[t]{\caption{Runs \label{tab:run}}
\begin{tabular}{cccrc}
\hline
$\beta$ & $r$ values & lattice & iupd & $\#$ of measurements \\
\hline
5.0 & $~2 - 8~$ & $36^3$ & 16000 & 1600 \\
    & $~7 - 9~$ & $40^3$ & 32000 & 3200 \\
    & $~8 - 12$ & $48^3$ & 48000 & 7000 \\
  &&&& \\
7.5 & $~4 - 8~$ & $48^3$ & 8000 & 1100 \\
    & $~7 - 12$ & $64^3$ & 18000 & 1100 \\
    & $11 - 16$ & $64^3$ & 36000 & 7200 \\
  &&&& \\
10.0 & $~2 - 7~$ & $48^3$ & 16000 & 2850 \\
     & $~6 - 9~$ & $48^3$ & 16000 & ~200 \\
     & $~8 - 14$ & $84^3$ & 24000 & 1100 \\
     & $13 - 19$ & $84^3$ & 36000 & 2250 \\
   &&&& \\
12.5 & $~2 - 9~$ & $48^3$ & 16000 & 2700 \\
     & $~8 - 14$ & $72^3$ & 24000 & 1150 \\
\hline
\end{tabular}
}
Another important parameter is the thickness of the time-slice over which the
sub-lattice averages are carried out.
We found that it was helpful to increase this thickness as
one goes from stronger to weaker coupling. We used values of two, four and six
as time-slice thicknesses.

To optimize the running 
time while keeping any finite volume effect under control we chose different lattice sizes
 for different ranges of $r$. Typically the smaller 
values of $r$ are much easier to get as they can be reliably obtained on smaller 
lattices and require much less sub-lattice updates. In principle of course results
 for all the 
different $r$ values could be obtained from the largest lattice but memory requirements 
prevent us from doing all the measurements in a single run. 

\TABLE{\caption{Lattice scales \label{tab:scales}}
\begin{tabular}{rllrll}
\hline
$\beta$ & $\frac{1}{2}\langle{\rm Tr}~u_P\rangle$& $~~a^2 \sigma$ & $r_0~~~~~~$ & $a~(fm)$ & $\quad\sigma r_0^2$ \\
\hline
5.0 & 0.786878 (7) & 0.097334 ~(6) & $3.9536 ~~(3)$ & 0.12648 ~(2) & 1.5214 ~(3) \\
7.5 & 0.861665 (4) & 0.038566 ~(6) & $6.2875 ~(10)$ & 0.07952 ~(1) & 1.5246 ~(7) \\
10.0 & 0.897683 (3) & 0.020606 ~(4) & $8.6022 ~~(8)$ & 0.05812 ~(1) & 1.5248 ~(4) \\
12.5 & 0.921100 (2) & 0.012742 (17) & $10.916 ~~~(3)$ & 0.04580 ~(1) & 1.5215 (29) \\
\hline
\end{tabular}
} 

The scale is set by the Sommer scale
$r_0=0.5~{\rm fm}$ which is implicitly defined
by $r_0^2 f(r_0)=1.65$, where $f(r)$ is the force between the static quark
and the anti-quark. We use our measured force values and
 interpolation to extract $r_0$. The values that we obtain are given in
table \ref{tab:scales}. We also see from table \ref{tab:scales} that $\sigma r_0^2$ is 
a constant $(\simeq 1.522)$ to a very good
approximation as expected in the scaling region close to the continuum limit.

\section{Results}
In this section
we present the results of our measurements. From the $\langle P^*P\rangle$
correlator one can extract the static quark-antiquark potential $V(r)$ by
\be
V(r)=-\frac{1}{T}\ln \langle P^*P(r)\rangle.
\ee
In principle the $q\bar q$ potential contains all the information about the flux tube, but 
it also contains an unphysical constant.
We therefore look directly at the first and the second 
derivative of this potential.

The first derivative of the potential gives us the force between the quark and the antiquark, 
while the second derivative gives us information about the subleading terms and how one approaches 
the asymptotic linearly 
rising behaviour of the $q\bar q$ potential. To facilitate our comparison with string models we 
actually compute a scaled second derivative which we call $c(\tilde r)$. This quantity should become the 
L\"uscher term ($=-\frac{(d-2)\pi}{24}$) asymptotically.

On the lattice these quantities are given by
\bea\label{force:eq}
F({\bar r})&=&V(r)-V(r-1) \\
c({\tilde r})&=&\frac{{\tilde r}^3}{2}[V(r+1)+V(r-1)-2V(r)]
\eea
where ${\bar r}=r+\frac{a}{2}+{\cal O}(a^2)$ and ${\tilde r}=r+{\cal O}(a^2)$ are defined as 
in \cite{lw2} to reduce lattice artifacts.

The theoretical predictions in continuum are given by string models (for large $r$) as well as 
perturbation theory (for small $r$). Since non-bosonic string models have been essentially ruled 
out \cite{LucTep}, we are going to concentrate on the potential due 
to the NG string, the so called Arvis potential \cite{Ar} given by
\be
V_{\rm Arvis}=\sigma r \left ( 1-\frac{(d-2)\pi}{12\sigma r^2}\right )^{1/2}.
\ee
Keeping in mind the results of 
L\"uscher and Weisz \cite{lweisz} and PS type effective string theories \cite{drum,pmd,unpub}
 we are going 
to compare our lattice 
data on force and $c(r)$ with leading order predictions to which all models reduce at sufficiently 
large $r$, to NLO expressions and expressions from the full Arvis potential. In three dimensions the 
expressions are given by 

\bea\label{eq:form}
{\bf L.O.}\quad  f(r)&=&\sigma + \left (\frac{\pi}{24}\right )\frac{1}{r^2} 
\qquad\qquad\qquad\qquad
~~~c(r)=-\frac{\pi}{24} \nonumber \\
{\bf N.L.O.}\quad f(r)&=&\sigma + \left (\frac{\pi}{24}\right )\frac{1}{r^2}+
\left (\frac{\pi}{24}\right )^2\frac{3}{2\sigma r^4} \qquad 
~c(r)=-\frac{\pi}{24}\left(1+\frac{\pi}{8\sigma r^2}\right ) \nonumber \\
{\rm Arvis}\quad f(r)&=&\sigma\left(1-\frac{\pi}{12\sigma r^2}\right )^{-1/2} \qquad\qquad 
\qquad c(r)=-\frac{\pi}{24}\left(1-\frac{\pi}{12\sigma r^2}\right)^{-\frac{3}{2}}.
\eea

The perturbative potential has been calculated at the one loop level by Schr\"oder \cite{pert}. 
He obtains 
\be
V_{\rm pert}(r)=\sigma_{\rm pert} r + \frac{g^2C_F}{2\pi}\ln g^2r + (\rm higher~order~terms)\\
\ee
with the perturbative string tension
$\sigma_{\rm pert}= \frac{7g^4C_FC_A}{64\pi}$. 
For $SU(2)$ $C_F=3/4$, $C_A=2$. 
The perturbative force and $c_{\rm pert}(r)$ can 
be computed by $f_{\rm pert}(r)=\frac{d V_{\rm pert}(r)}{d r}$ and $c_{\rm pert}(r)
=\frac{r^3}{2}\frac{d^2 V_{\rm pert}(r)}{d r^2}.$

\FIGURE{
\includegraphics[width=0.8\textwidth,angle=0]{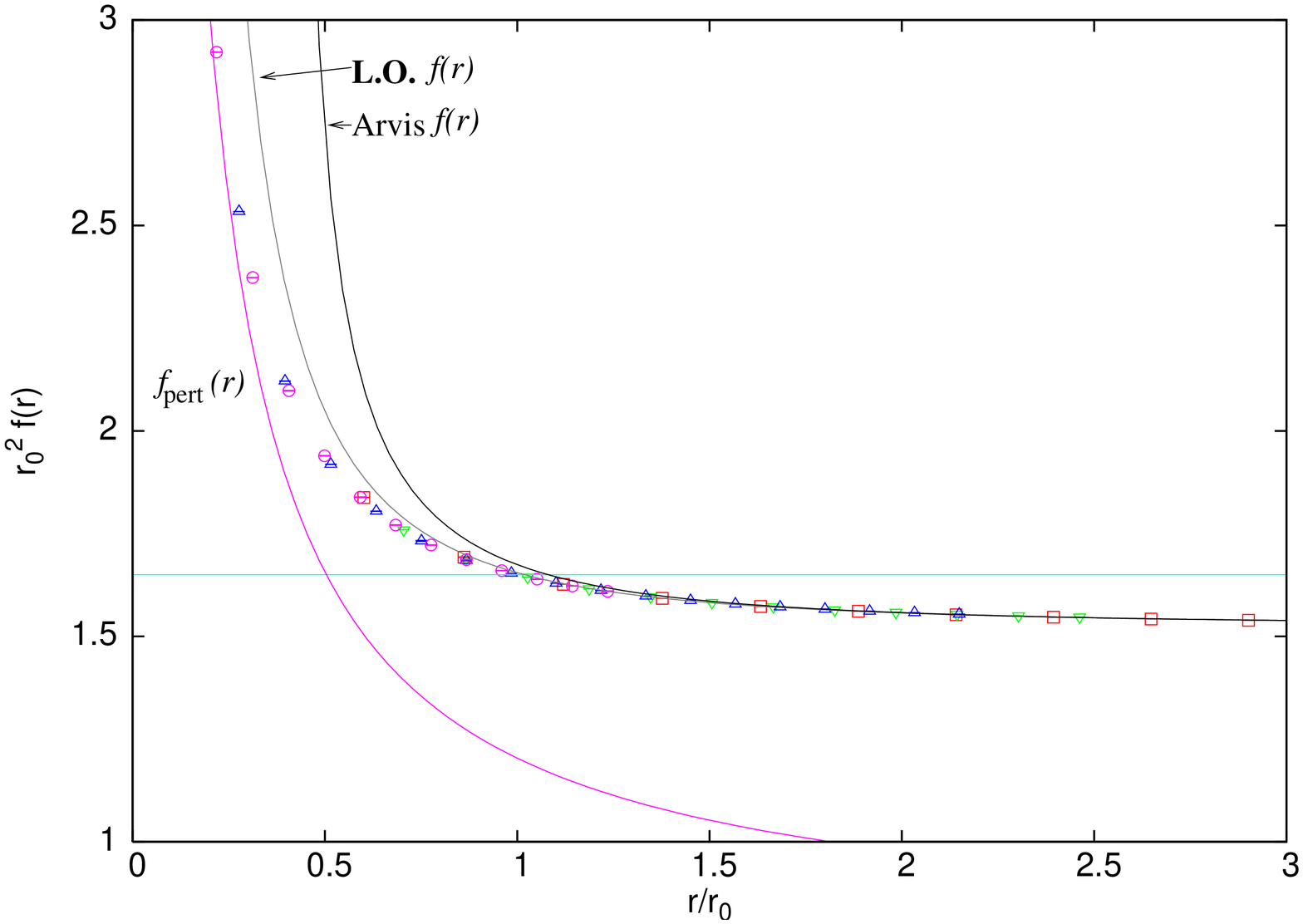}
\caption{$r_0^2f(r)$ vs $r/r_0$ in the $d=3$ $SU(2)$ case. The four different sets are
$\square ~(\beta=5)$, $\bigtriangledown ~(\beta=7.5)$, $\bigtriangleup ~(\beta=10)$ and 
$\bigcirc ~(\beta=12.5)$. Also shown is the 1-loop perturbation theory curve $f_{\rm pert}(r)$,
 as well as the leading order ({\bf L.O.} $f(r)$) and NG (Arvis $f(r)$) string predictions.
The horizontal line is $r_0^2f(r)=1.65$ and locates the Sommer scale.
 }
\label{force:fig}
}

We first look at the force data. 
Since we have four values of the coupling,
 we look at the continuum limit of the string tension. Following \cite{Teper} we too define 
$\beta_{\rm MF}=\beta\times\frac{1}{2}\langle{\rm Tr}~u_P\rangle$ and look at how $a\sqrt{\sigma}$ scales 
with $\beta_{\rm MF}$. We use $\beta_{\rm MF}=4/g^2a$ to convert the perturbative expressions 
to lattice units. Fitting the string tension to the form 
\be
a\sqrt{\sigma}=\frac{8}{\beta_{\rm MF}}\left (\frac{\sqrt{\sigma}}{g^2N}\right )_{\rm cont.}
+\frac{a_1}{\beta_{\rm MF}^2}+\frac{a_2}{\beta_{\rm MF}^3},
\ee
where $N$ refers to the gauge group $SU(N)$,
we obtain $\left (\frac{\sqrt{\sigma}}{g^2N}\right )_{\rm cont.}$, for $N=2$, 
 to be 0.16788(12) which 
is completely consistent with the values presented in \cite{Teper,Teper0}. 
As noted by Schr\"oder \cite{pert}, the full string tension is 1.47 times the perturbative
string tension and hardly changes for $N$ between 2 and 5. 
 We want to also mention that if we 
ignore the point $\beta=5$, then the data can be well described even without the coefficient 
$a_2$. In that case we obtain the continuum value to be 0.16736(10). Another point worth noting 
is that while we too find the coefficient $a_1$ to be negative, in agreement with \cite{Teper},
the coefficient $a_2$ is 
positive in our case. This is consistent with the higher $N$'s in \cite{Teper}, and may not be of much 
significance as in that work, $a_2$ for $SU(2)$ is consistent with 0 within 2$\sigma$.  

The values obtained for $\sigma r_0^2$ by fitting 
 the force to the Arvis form is quoted in table \ref{tab:scales}. The values given by the other 
two forms differs by less than 0.1\% from the quoted values. $\sigma r_0^2$
scales very nicely with the different beta values differing by less than 0.2\% from each other. 
 The fits were carried out in the range 2.4 - 2.9 $r_0$ for 
$\beta=5$ , 2.1 - 2.5 $r_0$ for $\beta=7.5$, 1.9 - 2.1 $r_0$ for $\beta=10$ and finally 
between 1.6 - 1.8 $r_0$ for $\beta=12.5$. The data for large $r$ at $\beta=12.5$ was taken 
from \cite{pushspec2}.

In Fig. \ref{force:fig} we plot 
$r_0^2f$ versus $r/r_0$ which is expected to become
a universal curve in the continuum limit. The horizontal line is $r^2f(r)=1.65$ and 
defines the Sommer scale $r_0$.
The grey line is the {\bf L.O.} result $\sigma r_0^2+\frac{\pi}{24}\left (\frac{r_0}{r}
\right )^2$ .
The black line is the Arvis curve and the magenta one the perturbative curve. The data 
starts departing from the one-loop perturbative curve around 0.22 $r_0$ or 1.8 GeV and 
joins onto the string curves around 1.5 $r_0$ or 260 MeV. 
The scaling exhibited by the data is very good with all the four different beta values 
falling on the same curve. This is mainly due to use of $\overline r$ instead of $r$ , as
 it eliminates lattice artefacts to a large extent. The value for 
$\sigma r_0^2$ for drawing the {\bf L.O.} and Arvis curves is taken from the fit at $\beta=5$
as it extends to the largest distance.
Beyond 1.5$r_0$ it is virtually 
impossible to distinguish the different theoretical curves as they are all dominated
 by the universal leading order behaviour with the string tension making up more than 95\% of the 
 force. The force data in fact gives the wrong impression of
the string description being good even at distances as small as $r_0$.

\FIGURE{
\includegraphics[width=0.8\textwidth,angle=0]{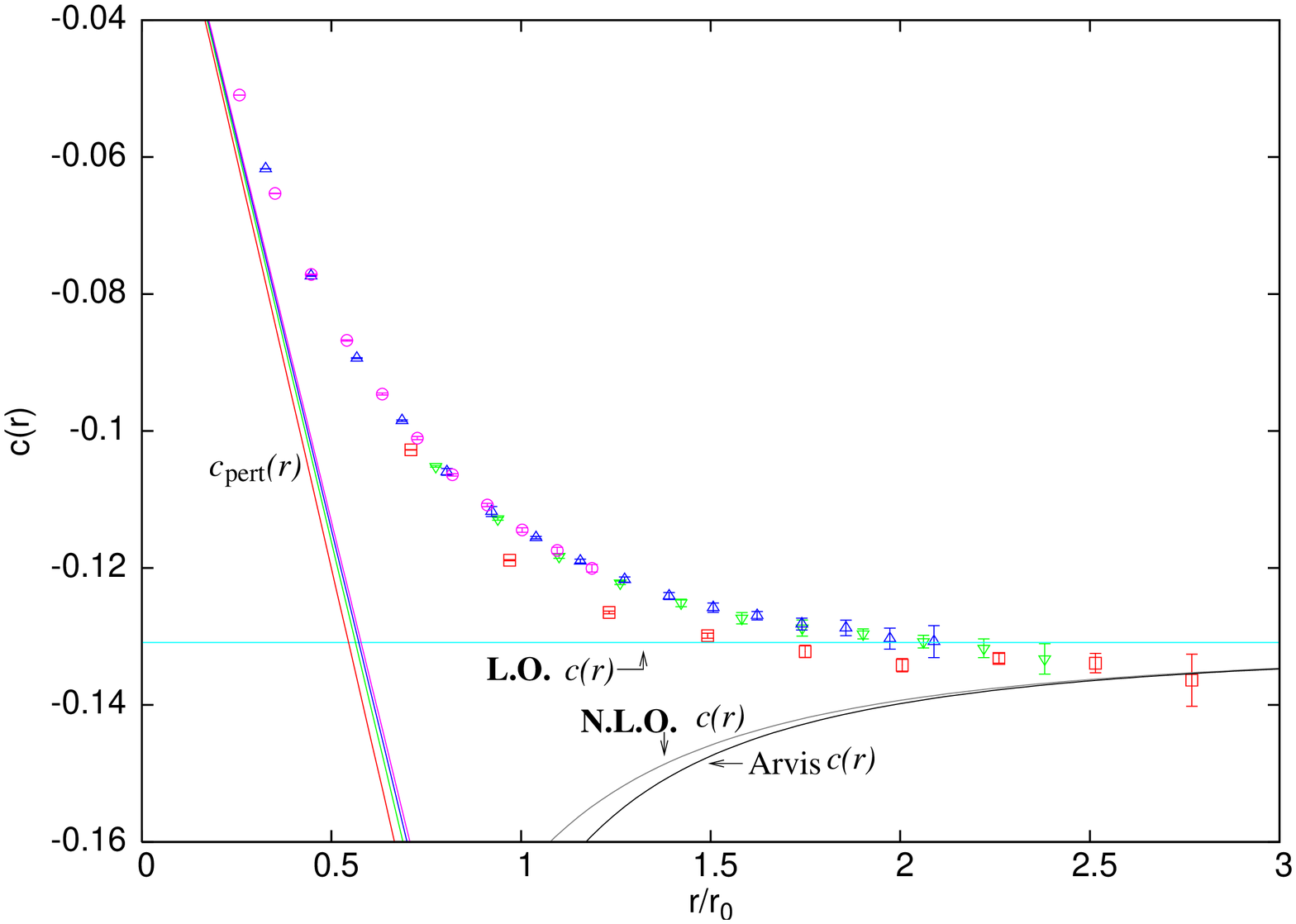}
\caption{$c(\tilde r)$ in $d=3$ $SU(2)$ case. The four different sets are
$\square ~(\beta=5)$, $\bigtriangledown ~(\beta=7.5)$, $\bigtriangleup ~(\beta=10)$ and
$\bigcirc ~(\beta=12.5)$. Also shown are 1-loop perturbation theory ($c_{\rm pert}(r)$), 
{\bf L.O., N.L.O.} and NG (Arvis) string model predictions.
 }
\label{fig:cr}
}

On the other hand, $c(\tilde r)$
 does not contain the string tension, and has a universal
value in the L.O. It is therefore more sensitive to the 
sub-leading behaviour of the flux tube. In Fig. 
\ref{fig:cr} we plot $c(\tilde r)$ (versus $r$ in units of $r_0$) for all four $\beta$ 
values along with the perturbative
curves as well as the three string model predictions given in eqn. \ref{eq:form}. The line 
and the symbol $\square$ in red corresponds to $\beta=5$, while the symbols $\bigtriangledown$, 
$\bigtriangleup$ and $\bigcirc$ in colours green, blue and magenta 
correspond to the $\beta$ values 7.5, 10 and 12.5 respectively. The blue horizontal line 
corresponds to the leading order prediction while the ash and black curves correspond to 
the NLO and Arvis forms respectively. 
We look at a wide range of $r$ starting from where the data almost touches the 
perturbative curves going all the way to the region where the string predictions hold.

The data seems to lie on top of each other exhibiting nice scaling behaviour  
as one goes to larger values of $r$. The $\beta=12.5$ and $\beta=10$ data lie on top of 
each other in the range 0.75 and 1.25 $r_0$ already. The $\beta=7.5$ set joins onto this 
at around 1.5$r_0$ and even the $\beta=5$ data joins up at around 2.25$r_0$. This points 
to the possibility that the continuum limit of the scale where the flux tube is well 
described by the Arvis curve can be obtained already on relatively coarse lattices. 
\section{Discussions}
In $d=3$, as also noted by Schr\"oder, infrared divergences prevent computation of the perturbative
potential beyond one loop. Infrared
counterterms have to be obtained non-perturbatively \footnote{NDH wishes to thank G. `t Hooft for an 
illuminating discussion on this.}. As our data goes down to distances of about 0.16 fermi, we can try 
to obtain these counter terms by looking at two loop terms in the perturbative potential. 
On dimensional grounds, one can expect the perturbative potential to be of the form
\be
V_{\rm pert}(r) = \frac{g^2C_F}{2\pi} \ln g^2 r + \frac{7C_FC_Ag^4}{64\pi} r + A g^4r \ln g^2 r + Bg^6 r^2
+\cdots
\ee
This gives
\be
c_{\rm pert}(r) = -\frac{g^2 r_0C_F}{4\pi} \frac{r}{r_0} +\frac{Ag^4r_0^2}{2}\left(\frac{r}{r_0}\right)^2 
+ Bg^6r_0^3\left(\frac{r}{r_0}\right)^3.
\ee
Since the first term is known, we determine $A$ and $B$ from the initial two points of our data
to be $A=0.013162(3)$ and $B=0.001089(1)$.
On our finest lattice we obtain $g^2r_0$, which is RG-invariant in the continuum limit, to be 
$3.792$. 
While $B$ contains effects of higher order terms, we expect the coefficient $A$ to be 
relatively well determined. 
From the ratio of the leading to the next to leading term in $V_{\rm pert}(r)$ as well as the data
on $c({\tilde r})$, we estimate the range of validity of first order 
perturbation theory to be about 
$ 0.1$ fermi (consistent with our estimate from the force data). 
Fig. \ref{LJ:fig} suggests that second order perturbation theory holds upto distances 
of about $0.2$ fermi.

To try to get an idea about the scale of string formation, we look at the 
 percentage of the total force carried by the string tension  
and the relative difference between the Arvis and the leading order force. From our 
data we find that the string tension constitutes 95\% of the force at around 1.3$r_0$,
 98\% at around 2.1$r_0$ and 99\% at around 2.9$r_0$. The relative difference, which
 gives us an idea about the importance of the subleading behaviour, is about 2\%
 at 1.02$r_0$, 1\% at 1.2$r_0$ and goes down to 0.1\% at about 1.9$r_0$. 

The type of string is even more difficult to identify.
At leading order, a variety of theories with different boundary conditions 
yield the universal L\"uscher term \cite{dietz}.
In fact all effective string theories of the
PS type and AdS/CFT correspondences \cite{naik} also yield this term.
The type is therefore determined by the sub-leading behaviour of the flux tube.
What can be clearly seen in the data is 
that the approach to the L\"uscher term is from below, consistent with effective bosonic string 
model predictions. At short distances the data matches perturbation theory.
Therefore 
 $c({\tilde r})$ crosses the asymptotic value at some intermediate point.
 For $SU(2)$, this intermediate point 
 seems to be around 2$r_0$ or 1 fermi. 
Here it is still not 
clear whether a string behaviour has set in. 
It is only at a still larger distance of about 2.75$r_0$, that the data
 seems to be well described by the Arvis curve. 

An interesting question is what happens to these scales for $SU(N)$ as $N$ increases.
It has been observed in \cite{HM5} that the intermediate $q\bar q$ separation at which 
$c(\tilde r)$
assumes its asymptotic value decreases with increasing $N$ at finite lattice spacing.
For $SU(3)$ the data on the coarsest lattice in ref. \cite{lw2} just about touches the 
Arvis curve at 1.8$r_0$. However both these scales shift towards larger $r$ as one 
approaches the continuum limit. 

Comparing the data with effective strings at higher orders may tell us if the scale 
of string formation is the one suggested by the Arvis curve or happens earlier. 
From ${\cal O}(r^{-4})$ and beyond, effective string theories motivate
parametrising the
leading deviations to $c(\tilde r)$ from the Arvis behaviour as
\be
\Delta c(\tilde r) = A\left(\frac{r_0}{r}\right)^4 + B\left(\frac{r_0}{r}\right)^6.
\ee
From our data in the
region $1.6-2.4~r_0$ we obtain, as best fit values,
$A=0.209(9), B=-0.235(24)$. The corresponding terms
of the Arvis potential are $A_{\rm Arvis}= -0.00725, B_{\rm Arvis}= -0.00145$.
It is of interest to know what the predictions of effective
string theories are for these.

\FIGURE{
\includegraphics[width=0.8\textwidth,angle=0]{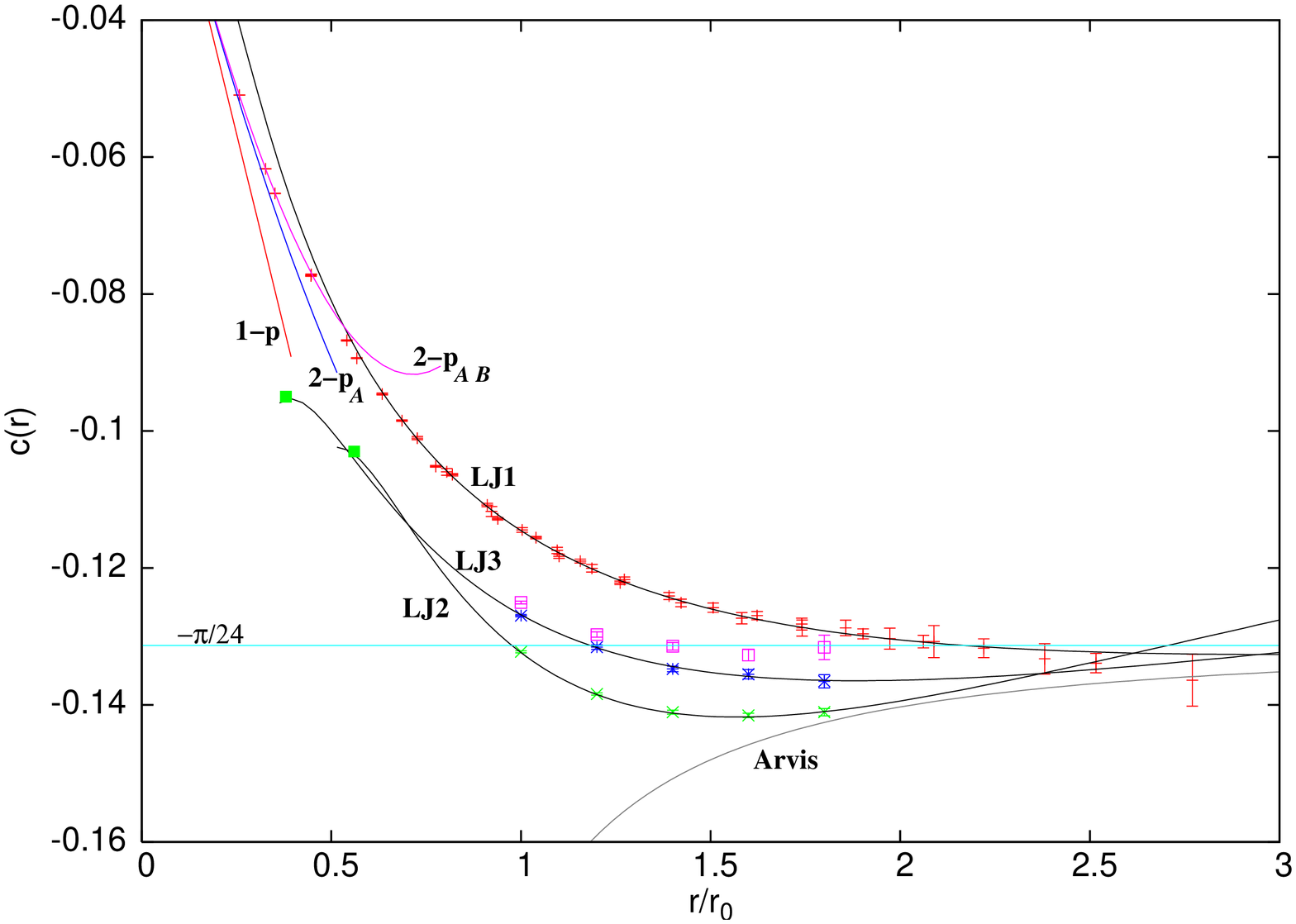}
\caption{ Curves of the type given in eqn. \ref{LJ} describing the $c(\tilde r)$ data 
in the intermediate region. The red + is the continuum limit $3-d\,SU(2)$ data. The green 
$\times$, the blue $\ast$ and the magenta $\square$ are $3-d\,SU(3)$ data from \cite{lw2}
with $r_0=3.30, 4.83$ and $6.71$ respectively.
The curves LJ1, LJ2 and LJ3 are given by $(a,b,n)$ of $(0.444, -0.258, 0.357)$ , 
$(0.458, -0.289, 0.691)$ and $(0.442, -0.287, 0.498)$ respectively. The perturbative curves marked 
1-p, 2-p$_A$ and 2-p$_{AB}$ are given by the first, the first two and all the three terms of 
the expression: 
$-\frac{g^2 r_0C_F}{4\pi} \frac{r}{r_0} +\frac{Ag^4r_0^2}{2}\left(\frac{r}{r_0}
\right)^2+Bg^6r_0^3\left(\frac{r}{r_0}\right)^3$ with $A=0.013162$ and $B=0.001089$.
 The green filled squares denote the smallest possible values of 
$r/r_0$ for which $c(\tilde r)$ can be calculated on that particular lattice.
 }
\label{LJ:fig}
}

At intermediate distances over a wide region of $r$ varying from $0.5r_0$ to $2.8r_0$, the data 
is very well described by a formula of the type 
\be\label{LJ}
c(\tilde r)= a\left (\frac{1}{x^{2n}}-\frac{1}{x^n}+\frac{b}{x^{3n}}\right )
\ee  with $a=0.444(4),~b=-0.258(2)
$ and $n=0.357(15)$. Existing $SU(3)$ data in 3-d \cite{lw2} also admit a similar description with 
nearly same values of $a$ and $b$, but with a different $n$. The curves are shown in Fig. 
\ref{LJ:fig}. At the moment it is not at all 
clear if there is any theoretical 
basis for such a description. However they certainly provide accurate interpolation formulae. 

\section{Conclusions}

In this article we have looked at the continuum limit behaviour of the $SU(2)$ flux 
tube at intermediate distances by measuring the static $q\bar q$ potential.
 Starting from a distance of about 0.1 fermi where the potential starts 
breaking away from 1-loop perturbation theory, we go to distances of about 1.4 fermi 
where the data begins to be well described by the Arvis potential. 

We look at the continuum limit of the string tension and find complete agreement 
between open and closed strings \cite{Teper}. 
Our data on $c(\tilde r)$ seems to approach the L\"uscher 
term from below as 
expected in bosonic string models. At distances below 0.15 fermi the data joins onto 
the perturbative values. A modified Lennard-Jones type empirical formula describes the 
data well in the intermediate region. 

Further directions of study include increasing the $q\bar q$ separation to confirm that 
the data indeed stays on the Arvis curve. For $SU(3)$ and $SU(5)$, it would be really interesting 
to push to the continuum limit to see if the behaviour seen in $SU(2)$ 
holds in those cases and find the distance where the data meets the Arvis curve. 

\acknowledgments
One of the authors, PM, gratefully acknowledges the numerous discussions 
 with Peter Weisz. NDH thanks the Hayama Centre for Advanced Studies for
hospitality. The simulations
were carried out on the teraflop Linux cluster KABRU at IMSc as part of the Xth plan project ILGTI.

\appendix
\TABLE[t]{\caption{Force and $c(\tilde r)$ ($\beta=5$)\label{tab5}}
\begin{tabular}{crlrl}
\hline
$r$ & $\overline{r}$ & $f(r)$ & $\tilde r$ & $c(\tilde r)$ \\
\hline
3 &  2.3790  & 0.11744 (1) & 2.808~ & $-$0.10274 (3) \\
4 &  3.4071  & 0.10816 (1) & 3.838~ & $-$0.11886 (8) \\
5 &  4.4322  & 0.10395 (1) & 4.876~ & $-$0.1265  (2) \\
6 &  5.4481  & 0.10177 (2) & 5.903~ & $-$0.1299  (4) \\
7 &  6.4579  & 0.10051 (2) & 6.920~ & $-$0.1322  (9) \\
8 &  7.4645  & 0.09974 (2) & 7.932~ & $-$0.1342 (10) \\
9 &  8.4692  & 0.09919 (1) & 8.941~ & $-$0.1330  (7) \\
10 & 9.4727  & 0.09883 (1) & 9.948~ & $-$0.1336 (16) \\
11 & 10.4755  & 0.09855 (1) &10.953~ & $-$0.1353 (41) \\
12 & 11.4777  & 0.09834 (2) &   $-$~~ &     $-$~~  \\
\hline
\end{tabular}}

\TABLE[t]{\caption{Force and $c(\tilde r)$ ($\beta=7.5$)\label{tab7.5}}
\begin{tabular}{crlrl}
\hline
$r$& $\overline{r}$ & $f(r)$ & $\tilde r$ & $c(\tilde r)$ \\
\hline
5&  4.4322 &  0.044497 (9)  & 4.876 &  $-$0.1052 (1)	 \\
6&  5.4481 &  0.042682 (11) & 5.903 &  $-$0.1128 (2)	 \\
7&  6.4579 &  0.041585 (12) & 6.920 &  $-$0.1183 (3)	 \\
8&  7.4645 &  0.040865 (8)  & 7.932 &  $-$0.1222 (3)	 \\
9&  8.4692 &  0.040375 (9)  & 8.941 &  $-$0.1251 (6)	 \\
10& 9.4727 &  0.040025 (10) & 9.948 &  $-$0.1273 (8)     \\
11& 10.4755 &  0.039767 (12) & 10.95 &  $-$0.1288 (11)    \\
12& 11.4777 &  0.039576 (5)  & 11.96 &  $-$0.1296 (7)	  \\
13& 12.4796 &  0.039424 (5)  & 12.96 &  $-$0.1308 (9)	  \\
14& 13.4812 &  0.039304 (6)  & 13.96 &  $-$0.1317 (14)    \\
15& 14.4825 &  0.039207 (6)  & 14.97 &  $-$0.1333 (22)    \\
16& 15.4837 &  0.039128 (7)  &   $-$ &       $-$        \\
\hline
\end{tabular}}

\TABLE[t]{\caption{Force and $c(\tilde r)$ ($\beta=10$)\label{tab10}}
\begin{tabular}{crlrl}
\hline
$r$ & $\overline{r}$ & $f(r)$ & $\tilde r$ & $c(\tilde r)$ \\
\hline
3&  2.3790 &  0.034245 (3) &  2.808 &  $-$0.06172 (1)  \\
4&  3.4071 &  0.028668 (4) &  3.838 &  $-$0.07737 (3) \\
5&  4.4322 &  0.025931 (5) &  4.876 &  $-$0.08937 (6) \\
6&  5.4481 &  0.024389 (6) &  5.903 &  $-$0.09849 (11)  \\
7&  6.4579 &  0.023412 (17)&  6.920 &  $-$0.10597 (49) \\
8&  7.4645 &  0.022772 (18)&  7.932 &  $-$0.11175 (72)\\
9&  8.4692 &  0.022346 (4) &  8.941 &  $-$0.11557 (18)\\
10 & 9.4727 &  0.022023 (5) &  9.948 & $-$0.11899 (28) \\
11& 10.4755 &  0.021781 (5) &  10.95 & $-$0.12170 (39) \\
12& 11.4777 &  0.021596 (5) &  11.96 & $-$0.12409 (51) \\
13& 12.4796 &  0.021450 (6) &  12.96 & $-$0.12580 (67) \\
14& 13.4812 &  0.021333 (4) &  13.96 &  $-$0.12696 (60) \\
15& 14.4825 &  0.021240 (4) &  14.97 &  $-$0.12819 (86) \\
16& 15.4837 &  0.021163 (5) &  15.97 &  $-$0.12875 (113) \\
17& 16.4847 &  0.021100 (5) &  16.97 &  $-$0.13032 (154) \\
18& 17.4856 &  0.021047 (5) &  17.97 &  $-$0.13076 (234) \\
19& 18.4864 &  0.021002 (6) &    $-$ &    $-$ \\
\hline
\end{tabular}
}

\TABLE[t]{\caption{Force and $c(\tilde r)$ ($\beta=12.5$)\label{tab12.5}}
\begin{tabular}{crlrl}
\hline
$r$& $\overline{r}$ & $f(r)$ & $\tilde r$ & $c(\tilde r)$ \\
\hline
3 & 2.3790 &  0.024521  (3) &  2.808 &  $-$0.05095 (1) \\	 
4 & 3.4071 &  0.019917  (4) &  3.838 &  $-$0.06530 (3) \\	 
5 & 4.4322 &  0.017607  (5) &  4.876 &  $-$0.07714 (5) \\	 
6 & 5.4481 &  0.016276  (5) &  5.903 &  $-$0.08677 (10) \\	 
7 & 6.4579 &  0.015432  (6) &  6.920 &  $-$0.09460 (15) \\	 
8 & 7.4645 &  0.014861  (7) &  7.932 &  $-$0.10105 (23) \\	 
9 &  8.4692 &  0.014453  (4)&  8.941 &  $-$0.10640 (17) \\
10&  9.4727 &  0.014155  (5)&  9.948 &  $-$0.11081 (25) \\
11& 10.4755 &  0.013930  (5)&  10.95 &  $-$0.11446 (35) \\
12& 11.4777 &  0.013756  (5)&  11.96 &  $-$0.11744 (47) \\
13& 12.4796 &  0.013618  (6)&  12.96 &  $-$0.12005 (55) \\
14& 13.4812 &  0.013508  (6)&    $-$ &    $-$ \\
\hline
\end{tabular}
}

\end{document}